\documentclass[10pt,conference]{IEEEtran}
\usepackage{amsmath,amssymb,amsthm}
\usepackage{graphicx}
\usepackage{booktabs}
\usepackage{xcolor}
\usepackage{cite}
\usepackage{soul}
\newcommand{\numSlots}{N}
\newcommand{\tcc}{\tau_{\mathrm{cc}}}
\newcommand{\tmem}{T_{\mathrm{mem}}}
\newcommand{\decay}{\lambda}
\newcommand{\cohSlots}{T}
\newcommand{\Fzero}{F_0}
\newcommand{\pgate}{p_{\mathrm{g}}}
\newcommand{\psrc}{q}
\newcommand{\Fthr}{F_{\mathrm{thr}}}

\newcommand{\pol}{\pi}
\newcommand{\rate}{R}
\newcommand{\numDel}{D}
\newcommand{\Opt}{Optimistic}
\newcommand{\Nest}{Nested}
\newcommand{\potl}{U}
\newcommand{\disc}{\gamma}
\newcommand{\numLayers}{L}
\newcommand{\emb}{h}
\newcommand{\score}{z}
\newcommand{\single}{g}
\newcommand{\pairW}{W}

\newcommand{\cost}{c}
\newcommand{\cstar}{c^{\star}}
\newcommand{\str}[1]{\mbox{\textrm{#1}}}

\newcommand{\E}{\mathbb{E}}

\def\paperversion{final}
\newif\ifshort
\newif\ifanonymous
\def\versionsubmission{submission}\def\versionfinal{final}\def\versionarxiv{arxiv}
\ifx\paperversion\versionsubmission \shorttrue \anonymoustrue
\else\ifx\paperversion\versionfinal \shorttrue \anonymousfalse
\else\ifx\paperversion\versionarxiv \shortfalse \anonymousfalse
\else \errmessage{Unknown version: write submission, final or arxiv in the definition of \string\paperversion}
\fi\fi\fi
\ifshort
  \newcommand{\arxivonly}[1]{\ignorespaces}
  \newcommand{\arxivfootnote}[1]{}
  \newcommand{\shortonly}[1]{#1}
\else
  \newcommand{\arxivonly}[1]{#1}
  \newcommand{\arxivfootnote}[1]{\footnote{#1}}
  \newcommand{\shortonly}[1]{}
\fi

\newcommand{\shorter}{\ifshort\looseness=-1\relax\fi}

\newtheorem{theorem}{Theorem}

\theoremstyle{definition}

\begin{document}
\bstctlcite{bstctl}

\title{Online Scheduling for Entanglement Distillation:\\ Learning When to Wait}

\ifanonymous
  \author{\IEEEauthorblockN{Anonymous Authors}}
\else
  \author{\IEEEauthorblockN{Panagiotis Promponas\IEEEauthorrefmark{1}, Spyros Tserkis\IEEEauthorrefmark{2}, and Leandros Tassiulas\IEEEauthorrefmark{1}}
          \IEEEauthorblockA{\IEEEauthorrefmark{1}Department of Electrical and Computer Engineering, Yale University, New Haven, CT, USA\\
          \IEEEauthorrefmark{2}Institute for Quantum Computing and Quantum Technologies, NCSR Demokritos, Greece\\
          \{panagiotis.promponas, leandros.tassiulas\}@yale.edu, spyrostserkis@gmail.com}}
\fi
\maketitle

\begin{abstract}
\shorter
Entanglement distillation probabilistically increases the fidelity of entangled pairs at the cost of reducing their number.
The outcome of each distillation operation takes time to communicate, during which the entangled pairs decohere while idle in their memories.
This raises a dilemma: wait for the outcome before proceeding, or continue optimistically at the risk of needlessly consuming further pairs.
We study online distillation scheduling, in which distillation trees are constructed dynamically to maximize the throughput of pairs that meet a required fidelity.
A policy decides which pairs to distill together and, for each operation, whether to wait for its outcome.
We derive an upper bound on the throughput of every scheduling policy and, by formulating the problem as a Markov decision process, train, with reinforcement learning, a policy that applies to any number of memories.
Compared with two baselines that always or never wait, the learned policy delivers as many pairs of the required fidelity as the better one when one of these two extremes is the best choice, and more otherwise.
\end{abstract}

\section{Introduction}

Quantum networks (QNs)~\cite{kimble2008,wehner2018} distribute entangled pairs to facilitate applications such as distributed quantum computing~\cite{cirac1999} and quantum cryptography~\cite{ekert1991}.
However, in practice, the entangled pairs shared between QN nodes are non-ideal, meaning that their fidelity to a Bell state is less than unity.
Entanglement distillation~\cite{bennett1996,deutsch1996,duer2007} is a probabilistic technique that consumes multiple imperfect pairs to produce fewer pairs of higher fidelity.
An important question is therefore not only
what fidelity distillation can achieve, but at what rate
it can deliver pairs above the fidelity an application requires, that is, its throughput.

\shorter
Both entanglement generation and distillation require classical information exchange to confirm success. In large scale QNs, these messages take far longer than the local operations~\cite{walter2017,yau2025}.
A conservative policy waits
for the outcomes of each distillation operation before
proceeding with its output. This avoids spending additional
pairs on intermediate results that may have failed, but
introduces communication delays during which stored pairs
decohere. This delay creates an opportunity for \emph{optimistic execution}~\cite{mobayenjarihani2023,hartmann2007}: the nodes may continue distilling an intermediate pair before learning whether the preceding distillation operation succeeded.
Optimistic execution reduces delay but risks the consumption of additional
pairs in operations whose earlier inputs have already failed.
The optimal policy might be neither to wait for every outcome nor to wait for none, but an intermediate one, in which some operations wait for their outcomes and others do not.
The question then becomes: \emph{how optimistic should
a distillation policy be?}

Distillation is usually described by a static circuit that combines a given collection of pairs.
In a continuously operating network, however, new elementary pairs
become available at random times, intermediate pairs compete
for limited quantum memories, and several distillation
trees may be constructed concurrently.
Each tree thus evolves over time and can be constructed online, by repeatedly choosing which stored pairs to distill together and which to discard.
Together with how optimistic to be, these choices form a \emph{distillation schedule}.
The optimal choice at any moment depends on which pairs are available, on their states, and on the trees that compete for them, so a schedule decided online can adapt to all of these.

\shorter
Existing work addresses parts of this problem.
Van Meter \textit{et al.}~\cite{vanmeter2009} showed for repeater chains that the rule for choosing which pairs to distill together changes the throughput by a factor of up to fifty, and proposed combining pairs of similar fidelity.
In entanglement routing, the tree of a static circuit has been optimized offline to maximize the number of pairs above a fidelity threshold~\cite{jia2024,mandilara2026}.
For a link with few memories, schemes that distill the stored pairs when a new pair is generated have been analyzed under decoherence~\cite{elsayed2024,inesta2026}, with bounds on the fidelity and the availability of a single stored pair over all such schemes~\cite{inesta2026}.
The capacity region of a quantum switch has been characterized for a given distillation protocol~\cite{panigrahy2023}.
Hartmann \textit{et al.}~\cite{hartmann2007} proposed applying distillation operations without waiting for the outcomes of the preceding ones.
Mobayenjarihani \textit{et al.}~\cite{mobayenjarihani2023} extend this to the messages that confirm the generation of a pair, and compare both with waiting for every outcome, under communication delay and memory decoherence, on fixed circuits for two or three memories per node.
Markov decision processes (MDPs) have been used to select distillation operations online, with every outcome known before the next decision~\cite{kumar2023,bhavsar2026}.
Closest to our approach, Yau \textit{et al.}~\cite{yau2025} train a policy with reinforcement learning (RL) to maximize the secret key rate under communication delay, for two or three memories per node. With three it may apply a second operation before the outcomes of the first arrive, and it is trained to deliver one pair from empty memories.
To our knowledge, no prior work determines where in a distillation tree to wait for the outcomes while trees are constructed continuously, and none bounds the throughput of every policy under memory decoherence and communication delay.

\shorter
Our main contributions are the following.
(i)~We formulate \emph{online distillation scheduling}, in which distillation trees are constructed dynamically to maximize the throughput of pairs that meet a required fidelity. As pairs are generated and decohere, a policy decides which pairs to distill together and whether to wait for the outcomes of each operation, so that the size, the shape and the optimism of every tree adapt to the current state of the link (Section~\ref{sec:model}).
(ii)~We derive an upper bound on the throughput of every scheduling policy under memory decoherence and communication delay, and show that it is attained in some settings (Section~\ref{sec:bound}).
(iii)~We formulate the problem as an MDP and train, with RL, a policy that applies to any number of memories and that each node can execute locally, without a central controller (Section~\ref{sec:mdp}).
(iv)~We compare the learned policy numerically with two baselines, one that always and one that never waits for the outcomes. It delivers as many pairs of the required fidelity as the better one when one of these extremes is the best choice, and more otherwise by constructing trees whose size and waiting points match neither baseline and depend on the state of the link (Section~\ref{sec:eval}).

\section{System Model}
\label{sec:model}

\subsection{Physical Model}
\label{sec:model-setting}

We consider two nodes connected by a link, each equipped with $\numSlots$ quantum memories, so that the nodes can share at most $\numSlots$ entangled pairs at a time.
Entangled pairs are generated over the link with fidelity $1/2 < \Fzero < 1$, whereas the application served by the nodes needs pairs of fidelity at least $\Fthr > \Fzero$, and the nodes close this gap by entanglement distillation.
Both generation and distillation are heralded by a classical message between the nodes: the heralding message of a generated pair, and the measurement outcomes of a distillation operation.
A message takes the one-way classical communication time $\tcc$ between the nodes, for example $0.1$\,ms over $20$\,km of optical fiber.
This delay is our time slot, as in~\cite{yau2025}: a message sent at the beginning of slot $t$ is read at the beginning of slot $t+1$.

\shorter
\textbf{Entanglement generation.}
A source in the middle of the link continuously emits entangled photon pairs, sending one photon of each pair to each node~\cite{jones2016}.
One memory at each node forms a \emph{memory pair}, which holds at most one entangled pair.
An empty memory pair is used for a generation attempt, in which each of its two memories can store an arriving photon.
The attempt succeeds if the two memories stored the two photons of the same entangled pair, which happens with probability $\psrc$, and we assume that attempts succeed independently of each other.
The success of an attempt cannot be verified locally, so the nodes exchange heralding messages.
A heralded pair is then an elementary pair of fidelity $\Fzero$ that can be distilled; a memory pair whose attempt failed is cleared and attempts again.
The probability $\psrc$ accounts for source emission, photon loss and storage in the memories; it depends on the entanglement generation protocol, and we treat it as a parameter.
We assume that the nodes can distill a pair only after its heralding message, even for $\psrc = 1$, when every attempt succeeds.

\textbf{Entanglement distillation.}
We consider two-to-one distillation with the DEJMPS protocol~\cite{deutsch1996}.
A \emph{distillation operation} consumes two stored pairs and, if it succeeds, leaves one pair of higher fidelity.
At each node, it applies one two-qubit gate to the two memories holding these pairs and measures one of them, which frees that memory at both nodes.
The operation has succeeded if the two measurement outcomes agree, which the nodes learn at the beginning of the next slot. 
A distilled pair can enter further operations, so the operations behind one delivered pair form a binary \emph{distillation tree}: its leaves are elementary pairs, and each internal node is one operation on two input pairs, elementary or distilled.
The tree thus specifies the distillation circuit of that pair.

\begin{figure}[t]
  \centering
  \includegraphics[scale=0.88]{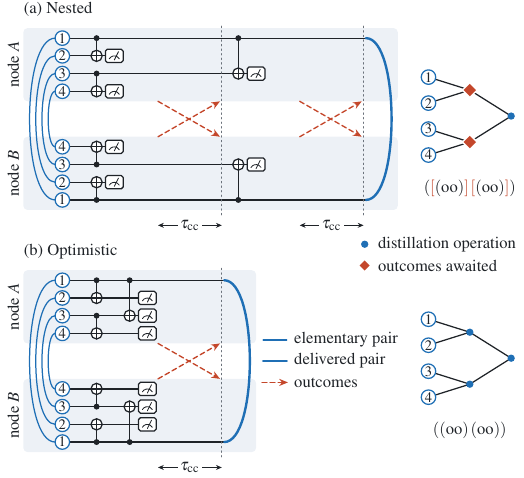}
  \caption{A distillation tree on four elementary pairs, divided into speculative blocks in two ways: (a)~the nested and (b)~the optimistic schedule. Each panel shows the circuit on the
  memories of nodes $A$ and $B$ (left) and the tree (right). Memories with the same number hold one
  elementary pair. Each distillation operation is drawn as a CNOT and a measurement at each node; its
  outcomes take $\tcc$ to cross the link and are known at the dotted line. Below
  each tree is its schedule denoted as a string.}
  \label{fig:protocol}
\end{figure}

The nodes need not wait for the outcomes of an operation before using its output in the next one.
Fig.~\ref{fig:protocol} shows a tree on four elementary pairs and its circuit, with and without waiting for the outcomes.
The tree has three operations: two at the first level and one at the second.
In (a), the \emph{nested} schedule, the second-level operation waits for the outcomes of both first-level operations; in (b), the \emph{optimistic} schedule, all three operations are applied within one slot, and the outcomes are exchanged at the end.
We call a group of operations that produce one pair without waiting for each other's outcomes a \emph{speculative block}.
In (a) each of the three operations is a block on its own, two in the first slot and one in the second; in (b) the three operations form one block.
In both, a pair survives only if every operation behind it succeeded.

We write a tree and its schedule as a string: $\str{o}$ denotes an elementary pair, $(xy)$ the pair that a distillation operation produces from the pairs $x$ and $y$, and $[x]$ the pair $x$ used only after the outcomes of its operations are known.
The nested and the optimistic schedule of Fig.~\ref{fig:protocol} read $([(\str{oo})][(\str{oo})])$ and $((\str{oo})(\str{oo}))$.
The two inputs of an operation need not be treated alike: in $([(\str{oo})](\str{oo}))$ the nodes wait for the outcomes of one first-level operation but not of the other.

\textbf{Noise model.}
Every pair in the model is Bell-diagonal, and we write its state as
$\rho = A\,|\Phi^{+}\rangle\langle\Phi^{+}| + B\,|\Phi^{-}\rangle\langle\Phi^{-}| + C\,|\Psi^{+}\rangle\langle\Psi^{+}| + D\,|\Psi^{-}\rangle\langle\Psi^{-}|$,
with $A + B + C + D = 1$; the fidelity of the pair is its overlap $A$ with the Bell state $|\Phi^{+}\rangle$.
An elementary pair is a Werner state, $A = \Fzero$ and $B = C = D = (1 - \Fzero)/3$.
The noise and the distillation below map Bell-diagonal states to Bell-diagonal states, so this description holds for every pair in memory.

\emph{Memory decoherence.}
We assume that a stored pair depolarizes: after $r$ slots in memory each of its components $x$ becomes
\begin{equation}
  \label{eq:decoherence}
  \tfrac14 + \decay^{r}\,\bigl(x - \tfrac14\bigr), \qquad \decay = e^{-\tcc/\tmem},
\end{equation}
where $\tmem$ is the coherence time of the memories, so a pair of fidelity $F$ has fidelity $\tfrac14 + \decay^{r}(F - \tfrac14)$ after $r$ slots.
The coherence time and the length of the link enter the model only through $\cohSlots = \tmem/\tcc$, the coherence time in slots.
The fidelity $\Fzero$ of an elementary pair is taken when its heralding message is read, after the slot its photon has spent in memory.

\shorter
\emph{Gate noise.}
We assume that each input pair of a distillation operation is depolarized with probability $\pgate$ before the operation, each component becoming $\tfrac14 + (1 - \pgate)(x - \tfrac14)$, which models the two-qubit gate error.
The measurements are assumed to be perfect.

\emph{Distillation operation.}
With these assumptions the outcome of an operation is as follows~\cite{duer2007}.
For input pairs with components $(A_1, B_1, C_1, D_1)$ and $(A_2, B_2, C_2, D_2)$ after the gate noise, the operation succeeds with probability
\begin{equation}
  \label{eq:dejmps-p}
  p = (A_1 + D_1)(A_2 + D_2) + (B_1 + C_1)(B_2 + C_2),
\end{equation}
and the output pair, conditional on success, has components
\begin{equation}
  \label{eq:dejmps-out}
  \begin{aligned}
    A' &= (A_1 A_2 + D_1 D_2)/p, & B' &= (A_1 D_2 + D_1 A_2)/p,\\
    C' &= (C_1 C_2 + B_1 B_2)/p, & D' &= (C_1 B_2 + B_1 C_2)/p.
  \end{aligned}
\end{equation}
\shorter
The map \eqref{eq:decoherence} applies for every slot a pair is stored, whether it waits for a partner or for the outcomes of an operation; \eqref{eq:dejmps-p} and \eqref{eq:dejmps-out} apply at each operation.
Inside a speculative block, the outcomes of an operation are not yet known when its output enters the next operation.
The components of this output are still given by~\eqref{eq:dejmps-out}, because the output of the block is used only if all its operations succeeded, which happens with probability equal to the product of~\eqref{eq:dejmps-p} over these operations.
A pair is delivered when the outcomes of its last operation arrive, one slot after the operation, so its delivered fidelity is $\tfrac14 + \decay\,(A' - \tfrac14)$.

\subsection{Problem Formulation}
\label{sec:model-problem}

\textbf{Scheduling and timing.}
The nodes continuously generate and distill entangled pairs,
reusing their memories as pairs are consumed or delivered.
Each slot has the following four parts.

\emph{Reading the messages.}
At the beginning of each slot, the nodes read the heralding
messages and measurement outcomes from the preceding slot.
Heralded elementary pairs can now be distilled, successful
distillation outputs can be used further or delivered, and
the memories of failed outputs are cleared.
We call the heralded elementary pairs and the successful distillation outputs \emph{available}.

\emph{Scheduling decisions.}
Based on the pairs now available, a \emph{scheduling policy}
decides, for each of them, whether to (i) deliver it, (ii) make it an input of a speculative block, or (iii) retain or discard it. Each speculative block specifies the order in which its inputs are combined, and the nodes wait for its outcomes after its last operation.
We assume that local quantum gates are much faster than
classical communication and treat them as instantaneous~\cite{yau2025},
so the nodes can execute all the selected blocks within
the current slot.\arxivfootnote{For example, two-qubit gates
last tens to hundreds of nanoseconds on superconducting
qubits~\cite{kjaergaard2020}, a few hundred nanoseconds
on neutral atoms~\cite{evered2023}, and as little as
$1.6\,\mu$s on trapped ions~\cite{schafer2018}.}
Since the intermediate measurement outcomes are not
needed to execute these blocks, the consumed qubits
can be measured together after the gates have been
applied. Thus, regardless of how many operations are
performed, only one round of measurements is needed.
Its duration is a fixed overhead, which we include in
the slot duration $\tcc$; for example, on superconducting
qubits, this overhead is negligible compared with the
communication delay~\cite{walter2017}.

\emph{Generation attempts.}
Every empty memory pair, including those freed by these measurements, deliveries, or discards, makes a new generation attempt.
Since the
source emits continuously, photons are already arriving
at the nodes, and we assume that storing them is fast
compared with $\tcc$.

\emph{Sending the messages.}
The nodes then send each other the heralding messages of these attempts and the measurement outcomes of the operations of this slot, which arrive at the beginning of the next slot.
While they travel, every stored pair undergoes one slot
of decoherence according to \eqref{eq:decoherence}.

\begin{figure}[t]
  \centering
  \includegraphics[scale=0.88]{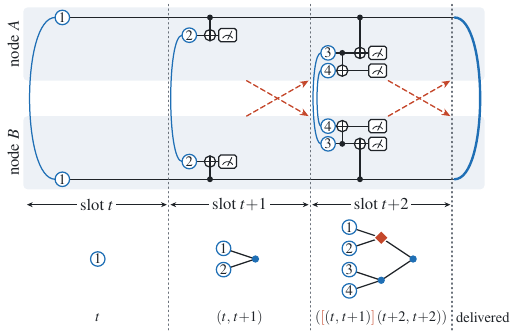}
  \caption{Dynamic view of the running system: one distillation tree as it is constructed over time. Time runs from left to right in slots. Below each slot are the part of the tree constructed so far and its string, where each ``$\str{o}$'' of Fig.~\ref{fig:protocol} is replaced by the slot in which that elementary pair becomes available, so that the first set $(\str{oo})$ becomes, for example, $(t,\,t+1)$. The policy constructs other trees at the same time on the remaining memories; these are not drawn.}
  \label{fig:schedule}
\end{figure}

\textbf{Online construction of trees.}
In Fig.~\ref{fig:protocol}, all the elementary pairs are available at the start.
However, the running system is dynamic: they may become available in different slots as the policy constructs a tree.
\arxivonly{\par
To capture the schedule of a tree in the running system, the tree string notation of Fig.~\ref{fig:protocol} must also record when the elementary pairs become available and when each speculative block is executed.
We replace each ``$\str{o}$'' with the slot in which that elementary pair becomes available, which suffices because all elementary pairs start in the same state.
An operation $(t_1,\,t_2)$ with $t_1 \le t_2$ is then applied in slot~$t_2$, and $[(t_1,\,t_2)]$ is available in slot~$t_2+1$, when the outcomes arrive, as if it were an elementary pair heralded in that slot.
In the highly unlikely case that a schedule delays the operation by $w$ slots after both inputs are available, we write $(t_1,\,t_2)^w$.
The string thus determines when every block is executed.}%
\shortonly{To capture the schedule of a tree in the running system, we replace each ``$\str{o}$'' with the slot in which that elementary pair becomes available. We write $(t_1,\,t_2)^w$ for an operation delayed by $w$ slots after both inputs are available, which a schedule has no reason to do.}
Fig.~\ref{fig:schedule} gives an example when elementary pairs are generated stochastically.

\shorter
\textbf{Objective.}
Our goal is to find a scheduling policy that maximizes the
\emph{throughput}, the long-run rate of delivering entangled
pairs with fidelity at least $\Fthr$. Let $\numDel_t^{\pol}$
denote the number of such pairs delivered in slot $t$ under
policy $\pol$.
Let $\Pi$ be the set of causal scheduling policies, which make the scheduling decisions of each slot using only the messages read up to that slot.
The optimal throughput, in pairs per second, is
\begin{equation}
\rate^\star(\Fthr)
=
\sup_{\pol\in\Pi}
\liminf_{H\to\infty}
\frac{1}{H\tcc}
\E_{\pol}\left[
\sum_{t=0}^{H-1} \numDel_t^{\pol}
\right].
\label{eq:optimal-throughput}
\end{equation}

\section{Throughput Bound}
\label{sec:bound}
The optimal throughput~\eqref{eq:optimal-throughput} is hard to compute, since the supremum ranges over the infinite set of causal policies, so we cannot tell how far a given policy is from optimal.
We therefore bound it from above, and use the bound in Section~\ref{sec:eval} as the reference for the policies of Section~\ref{sec:mdp}.
To obtain the bound, we assign each pair a \emph{cost}, which counts the memory pairs used to produce it, each multiplied by the number of slots for which it is used.

We define the cost of an elementary pair as $1/\psrc$.
It takes $1/\psrc$ generation attempts on average, each of which uses a memory pair for one slot.
For a pair $x$ produced by a speculative block with inputs $x_1,\dots,x_m$, where $x_i$ was retained for $r_i$ slots after it became available, we define
\begin{equation}
  \cost(x) = \frac{1 + \sum_{i=1}^{m} \bigl(\cost(x_i) + r_i\bigr)}{p(x)}.
  \label{eq:cost}
\end{equation}
Here $p(x)$ is the probability that all $m-1$ operations of the block succeed, computed with \eqref{eq:dejmps-p} and \eqref{eq:dejmps-out} from the states of the inputs.
The $1$ counts the memory pair that holds $x$ while the outcomes of the block cross the link, and the division by $p(x)$ counts the $1/p(x)$ executions that the block needs on average to succeed.
Since all elementary pairs start in the same state, the tree and schedule of $x$ determine $p(x)$, $\cost(x)$ and the fidelity $F(x)$ of $x$ when it becomes available.

Let $\cstar(\Fthr)$ be the smallest cost $\cost(x)$, over all schedules and generation outcomes, of a pair $x$ with $F(x) \ge \Fthr$, and let $\cstar(\Fthr) = \infty$ if no such pair exists.

\begin{theorem}
\label{thm:bound}
If $\Fthr > 1/4$, the optimal throughput satisfies $\rate^\star(\Fthr) \le \numSlots/\bigl(\cstar(\Fthr)\,\tcc\bigr)$.
\end{theorem}

\begin{proof}
Fix a causal policy $\pol$.
Let $\mathcal{P}_t$ be the set of pairs available for the scheduling decisions of slot~$t$.
For $x \in \mathcal{P}_t$, let $\cost_t(x) = \cost(x) + r$, where $r = t - s$ is the number of slots for which $x$ has been retained since it became available in slot~$s$.
If $x$ enters a block in slot~$t$, $\cost_t(x)$ is its term in the numerator of~\eqref{eq:cost}.
Let $\Phi_t = \sum_{x \in \mathcal{P}_t} \cost_t(x)$; $\Phi_0 = 0$, since the memory pairs are empty at the start.
Let $\mathcal{G}_t$ be everything that has happened up to and including the scheduling decisions of slot~$t$, and let $\mathcal{L}_t \subseteq \mathcal{P}_t$ be the set of pairs delivered or discarded in slot~$t$.
We first show that
\begin{equation}
  \E[\Phi_{t+1} \mid \mathcal{G}_t] = \Phi_t + \numSlots - \sum_{x \in \mathcal{L}_t} \cost_t(x).
  \label{eq:potential}
\end{equation}

In slot~$t$, the policy delivers or discards the pairs of $\mathcal{L}_t$, retains a set $\mathcal{R}_t$ of pairs, and uses every other pair of $\mathcal{P}_t$ as an input of a speculative block.
Let $\mathcal{B}_t$ be the set of outputs of these blocks, and $I(y)$ the set of inputs of the block with output $y$.
Since every pair of $\mathcal{P}_t$ belongs to exactly one of these groups,
$\Phi_t = \sum_{x \in \mathcal{L}_t} \cost_t(x) + \sum_{x \in \mathcal{R}_t} \cost_t(x) + \sum_{y \in \mathcal{B}_t} \sum_{x \in I(y)} \cost_t(x)$.
Let $\mathcal{A}_t$ be the set of memory pairs that make a generation attempt in slot~$t$.
Since every memory pair that holds neither a pair of $\mathcal{R}_t$ nor an output of $\mathcal{B}_t$ makes an attempt, ${|\mathcal{R}_t| + |\mathcal{B}_t| + |\mathcal{A}_t| = \numSlots}$.
A pair $x \in \mathcal{R}_t$ is still available in slot~$t+1$, with $\cost_{t+1}(x) = \cost_t(x) + 1$.
An output $y \in \mathcal{B}_t$ becomes available in slot~$t+1$ only if its block succeeds, and then ${\cost_{t+1}(y) = \bigl(1 + \sum_{x \in I(y)} \cost_t(x)\bigr)/p(y)}$ by~\eqref{eq:cost}.
The pair generated by a memory pair $a \in \mathcal{A}_t$ becomes available in slot~$t+1$ only if the attempt succeeds, and then costs $1/\psrc$.
With $\mathbf{1}_y$ and $\mathbf{1}_a$ the indicators of these successes,
$\Phi_{t+1} = \sum_{x \in \mathcal{R}_t} \bigl(\cost_t(x) + 1\bigr) + \sum_{a \in \mathcal{A}_t} \mathbf{1}_a/\psrc + \sum_{y \in \mathcal{B}_t} \bigl(\mathbf{1}_y/p(y)\bigr) \bigl(1 + \sum_{x \in I(y)} \cost_t(x)\bigr)$.
Given $\mathcal{G}_t$, the block with output $y$ succeeds with probability $p(y)$ and an attempt with probability $\psrc$, so
$\E[\Phi_{t+1} \mid \mathcal{G}_t] = \sum_{x \in \mathcal{R}_t} \bigl(\cost_t(x) + 1\bigr) + |\mathcal{A}_t| + \sum_{y \in \mathcal{B}_t} \bigl(1 + \sum_{x \in I(y)} \cost_t(x)\bigr)$,
and subtracting the expression for $\Phi_t$ gives~\eqref{eq:potential}.

Taking expectations in~\eqref{eq:potential} and summing over $t = 0, \dots, H-1$ gives
\ifshort
$\E[\Phi_H] = \numSlots H - \E\bigl[\sum_{t=0}^{H-1} \sum_{x \in \mathcal{L}_t} \cost_t(x)\bigr]$.
\else
\begin{equation}
  \E[\Phi_H] = \numSlots H - \E\Bigl[\sum_{t=0}^{H-1} \sum_{x \in \mathcal{L}_t} \cost_t(x)\Bigr].
  \label{eq:summed}
\end{equation}
\fi
Since $\Phi_H \ge 0$ and all costs are nonnegative, the pairs delivered in the first $H$ slots have an expected total cost of at most $\numSlots H$.

Every pair $x$ delivered in slot~$t$ with fidelity at least $\Fthr$ has $\cost_t(x) \ge \cstar(\Fthr)$.
By~\eqref{eq:decoherence}, the fidelity of a retained pair moves toward $\tfrac14$: if $x$ was retained for $r$ slots, its fidelity at delivery, $\tfrac14 + \decay^{r}\bigl(F(x) - \tfrac14\bigr)$, lies between $\tfrac14$ and the fidelity $F(x)$ that $x$ had when it became available.
Since the fidelity at delivery is at least $\Fthr > \tfrac14$, we have $F(x) \ge \Fthr$, and the definition of $\cstar(\Fthr)$ gives $\cost_t(x) \ge \cost(x) \ge \cstar(\Fthr)$.
Hence the expected number $\E\bigl[\sum_{t=0}^{H-1} \numDel_t^{\pol}\bigr]$ of pairs delivered with fidelity at least $\Fthr$ in the first $H$ slots is at most $\numSlots H/\cstar(\Fthr)$.
Dividing by $H\tcc$ and letting $H \to \infty$ shows that the throughput of $\pol$ is at most $\numSlots/\bigl(\cstar(\Fthr)\,\tcc\bigr)$.
Since $\pol$ is arbitrary, the same holds for $\rate^\star(\Fthr)$.
\end{proof}

\textbf{Tightness.}
Theorem~\ref{thm:bound} holds for every policy, including those that construct many trees at once, yet its bound depends only on the cheapest schedule of a single tree under the most favorable generation outcomes.
\arxivonly{The proof obtains the bound from~\eqref{eq:summed} by dropping three nonnegative terms: $\Phi_H$, the costs of the pairs discarded or delivered with fidelity below $\Fthr$, and the excess $\cost_t(x) - \cstar(\Fthr)$ of every pair $x$ delivered in slot~$t$ with fidelity at least $\Fthr$.
To attain the bound, it suffices that these terms stay bounded as $H \to \infty$.
This excess is zero if $x$ is produced by a cheapest schedule under the most favorable generation outcomes, in which no pair waits.
For $\psrc < 1$, the slots in which elementary pairs become available are random, so such outcomes occur only by chance.}
The bound is tight, for example, if $\psrc = 1$, the cheapest way to reach $\Fthr$ is a single speculative block on $n$ elementary pairs, and $n+1$ divides $\numSlots$.
The policy that splits the memory pairs into groups of $n+1$ attains it. In every slot, each group applies the block to $n$ freshly heralded pairs and delivers the output of its previous block, held in the remaining memory pair, if that block succeeded, so no pair waits and no memory pair is idle.
Each group delivers one pair per slot with the success probability $p$ of the block, so the throughput is $\numSlots p/\bigl((n+1)\,\tcc\bigr)$, which equals the bound since $\cstar(\Fthr) = (n+1)/p$ by~\eqref{eq:cost}.

\section{Scheduling Policies}
\label{sec:mdp}

\subsection{Baseline Policies}
\label{sec:baselines}

\shorter
The nested and the optimistic schedule of Fig.~\ref{fig:protocol} are the two extremes.
In the nested schedule, every operation waits for the outcomes of the operations that produced its inputs, so each operation is a speculative block of its own.
In the optimistic schedule, no operation waits for outcomes, so the whole tree is a single speculative block.
We use them as baselines, on the balanced tree with $d$ levels and $n = 2^d \le \numSlots$ elementary pairs.
Fig.~\ref{fig:protocol} shows the case $n = 4$.
In the running system, the elementary pairs of a tree become available in different slots, and those that are retained decohere.
Each baseline therefore also specifies which of the available pairs it combines and a fidelity cutoff, set by a parameter $\alpha \in (0,1)$, for the pairs it retains.
It delivers every pair that its tree produces, so $n$ and $\alpha$ must be chosen for the given $\Fthr$, as we do in Section~\ref{sec:eval}.

\textbf{\Opt$(n,\alpha)$.}
Under the optimistic schedule, only elementary pairs are ever retained.
While at least $n$ elementary pairs are available, the policy takes the $n$ that became available first, places them on the leaves of the tree in that order and applies its $n-1$ operations.
The output of the block is delivered in the next slot if the block succeeded.
The remaining pairs, fewer than $n$, are retained.
A pair retained for $r$ slots has fidelity $\tfrac14 + \decay^{r}(\Fzero - \tfrac14)$ by~\eqref{eq:decoherence}, and the policy discards it once this fidelity is below $\alpha\Fzero$, that is, after a fixed number of slots.

\textbf{\Nest$(n,\alpha)$.}
Under the nested schedule, distilled pairs can also be retained while they wait for a partner, so the policy must decide which pairs to combine.
It assigns each pair a level: an elementary pair has level $0$, and an operation on two pairs of level $k-1$ produces a pair of level $k$.
In every slot and for each $k < d$, it combines the available pairs of level $k$ two by two, in the order in which they became available.
A pair of level $d$ is delivered as soon as it becomes available; a pair without a partner of its level is retained.
While a pair is retained for a partner of its level, it decoheres, and its fidelity can fall to that of a pair of a lower level.
Let $F_k$ denote the fidelity with which a pair of level $k$ becomes available if no pair in its tree was retained.
When the fidelity of a pair of level $k$, with $1 \le k < d$, is below $\alpha F_{k-1}$, the policy reduces its level to $k-1$ and combines it with pairs of that level.
A pair whose fidelity is below $\alpha\Fzero$ is discarded.
A pair delivered after such a reduction has more than $n$ elementary pairs in its tree.
In Section~\ref{sec:eval-required} we extend \Nest$(n,\alpha)$ so that every pair it delivers meets a threshold $\Fthr$: a pair of level $d$ with fidelity below $\Fthr$ is distilled further instead of being delivered.

\subsection{Learned Policy: MDP Formulation}
\label{sec:learned}

\shorter
As defined above, the baselines decide two things in advance that a policy can instead decide in every slot, according to the stored pairs.
The first is which operations wait for the outcomes of the operations that produced their inputs: all of them under \Nest$(n,\alpha)$, and none under \Opt$(n,\alpha)$.
In the schedule of Fig.~\ref{fig:schedule}, the nodes wait for the outcomes of the operation on pairs 1 and 2, which arrive in the slot in which pairs 3 and 4 become available, but not for the outcomes of the operation on pairs 3 and 4. Hence neither baseline constructs this schedule.
The second is the size of a tree, which is set by $n$, whereas the number of elementary pairs that a tree needs to reach $\Fthr$ depends on how long its pairs have been retained, which is random.
We therefore train a policy that fixes neither: in every slot it chooses the speculative blocks to execute and the pairs to discard according to the stored pairs.
To this end, we formulate the problem as an MDP.

\shorter
\emph{State.}
The state describes each of the $\numSlots$ memory pairs when an action is taken.
An empty memory pair is marked as such.
For a memory pair that holds an available pair, the state contains the components $(A, B, C, D)$ of that pair, calculated with the noise model of Section~\ref{sec:model-setting}.
For a memory pair that holds the output of a speculative block executed in the current slot, whose outcomes are not yet known, it contains the probability that all operations of the block succeed and the components that the output has if they do, calculated in the same way.
For every stored pair, the state also contains the number of slots for which the pair has been stored, and for the output of a block the number of operations that the block contains so far.

\emph{Actions.}
The policy takes the scheduling decisions of a slot as a sequence of actions.
An action (i)~applies an operation to two stored pairs, which starts or continues a speculative block, (ii)~discards an available pair, or (iii)~ends the decisions of the slot.
The operation starts a block if both pairs are available.
It continues a block if one of them is the output of an operation of the same slot, and joins two blocks into one if both are.
When the policy ends the decisions of the slot, the nodes exchange the outcomes of all its operations.
Deliveries are not actions: when the messages are read, every available pair with fidelity at least $\Fthr$ is delivered, and a pair below $\Fthr$ remains stored for further operations.

\emph{Reward.}
The action that ends the decisions of slot~$t$ receives the reward $\numDel_{t+1}^{\pol}$, the number of pairs delivered when the messages of slot~$t$ are read, and every other action receives zero.
The long-run average reward per slot, divided by $\tcc$, is thus the throughput of the policy, which~\eqref{eq:optimal-throughput} maximizes.
With this reward alone, however, an operation is rewarded only when a delivery follows, possibly several slots later.
During training we therefore add a potential-based shaping term~\cite{ng1999} to the reward of every action.
The potential $\potl_\ell$ is the highest fidelity in the state in which the policy takes its $\ell$-th action, where the output of a block counts with the fidelity that it has if the block succeeds.
If no pair is stored, $\potl_\ell = 0$.
The shaping term of the $\ell$-th action is $0.5\,(\disc\,\potl_{\ell+1} - \potl_\ell)$, where $\disc$ is the discount factor of the training, so an operation that raises the highest fidelity is rewarded at once.

\subsection{Learned Policy: Architecture and Training}
\label{sec:network}

\begin{figure}[t]
  \centering
  \includegraphics[scale=0.836]{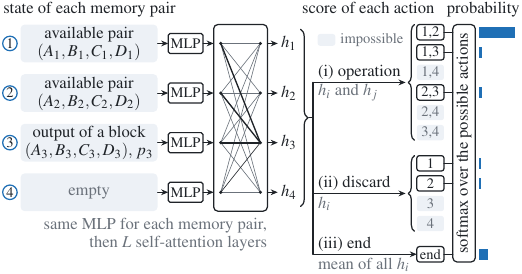}
  \caption{The policy network, drawn for $\numSlots = 4$ memory pairs in an example state. Each action receives a score, computed from the embeddings written under its arrow, and a softmax over the scores of the possible actions gives the probability of each action. The bars illustrate these probabilities.}
  \label{fig:network}
\end{figure}

Fig.~\ref{fig:network} shows the network that computes the policy.

\shorter
\emph{Encoder.}
The input of the encoder is one vector per memory pair, which contains the state of that memory pair.
Two properties of the problem determine how the encoder processes these vectors.
First, the memory pairs are interchangeable, so a decision must not depend on how they are numbered.
The encoder therefore applies the same multilayer perceptron (MLP) to the vector of every memory pair, and the number of a memory pair is not part of the input.
Second, whether a pair should enter an operation, be retained, or be discarded depends on the other stored pairs.
For this reason the outputs of the MLP pass through $\numLayers$ self-attention layers, which have the structure of the encoder layers of~\cite{vaswani2017}: multi-head self-attention followed by a feed-forward network.
The self-attention updates the vector of each memory pair with a weighted combination of the vectors of all memory pairs, where the weights are computed from the vectors themselves.
The output of the last layer is the \emph{embedding} $\emb_i$ of memory pair~$i$.

\shorter
\emph{Action scores.}
The network assigns a score $\score$ to every action, and the three kinds of action use the embeddings differently, as marked in Fig.~\ref{fig:network}.
(i)~An operation has two inputs, so its score $\score_{ij}$ is computed from the embeddings $\emb_i$ and $\emb_j$ of their memory pairs.
We build this score from two terms: one that depends on a single memory pair, and one that depends on both.
The first term, $\single_i$, is computed from $\emb_i$ by a linear layer and expresses how suitable the pair in memory pair~$i$ is for an operation, whatever the other input.
The second term, $\emb_i^{\top} \pairW \emb_j$ with a trained matrix $\pairW$, expresses how suitable the pair in memory pair~$j$ is as the other input of that operation.
The sum $\single_i + \emb_i^{\top} \pairW \emb_j$ thus scores the operation as the choice of the pair in memory pair~$i$, followed by the choice of the pair in memory pair~$j$ as its partner; $\single_j + \emb_j^{\top} \pairW \emb_i$ scores the same operation with the two choices in the other order.
Since an operation treats its two inputs alike, the score combines both orders:
$\score_{ij} = \log\bigl(e^{\,\single_i + \emb_i^{\top} \pairW \emb_j} + e^{\,\single_j + \emb_j^{\top} \pairW \emb_i}\bigr)$.
(ii)~A discard concerns one memory pair, so its score is computed from $\emb_i$ alone, by a second linear layer.
(iii)~Ending the decisions of the slot concerns no particular memory pair, so its score is computed from the mean of all embeddings, by a third linear layer.
Not every action is possible in every state: an operation needs a pair in both memory pairs and a discard needs an available pair, whereas the decisions of the slot can always be ended.
The policy takes each possible action with probability proportional to $e^{\score}$, which is the softmax over the possible actions in Fig.~\ref{fig:network}.

\emph{Training.}
We train the policy with proximal policy optimization (PPO)~\cite{schulman2017}, an actor-critic method.
In our network the actor is the part drawn in Fig.~\ref{fig:network}, and the critic is a second MLP that computes its estimate of the value of a state from the mean of the embeddings, so the two share the encoder.
The critic is used only during training.
We train one policy for each setting of the link parameters and each $\Fthr$.

\emph{Remarks.}
The learned policy has \ifshort two\else four\fi{} properties that matter in practice.
First, no parameter of the network depends on $\numSlots$, so a policy trained for one $\numSlots$ can be applied to a link with another; only the computation of an action grows with $\numSlots$, as $\numSlots^2$.
Second, the policy needs no central controller.
At the beginning of a slot both nodes know the state, which is determined by the messages they have exchanged, and once trained the policy takes in every state the possible action with the highest score.
Each node can thus run its own copy of the network, and both obtain the same actions.
\arxivonly{Third, instead of training one policy for each setting and each $\Fthr$, the encoder can be given $\Fthr$ and the parameters of the link as additional inputs, so that a single policy is trained for a range of settings. We leave this for future work. Fourth, the reward can be changed to train the policy for an objective other than the throughput, according to what the network needs; the learned policy of~\cite{yau2025}, for example, maximizes the secret key rate.}

\begin{figure*}[t]
  \centering
  \includegraphics[scale=0.92]{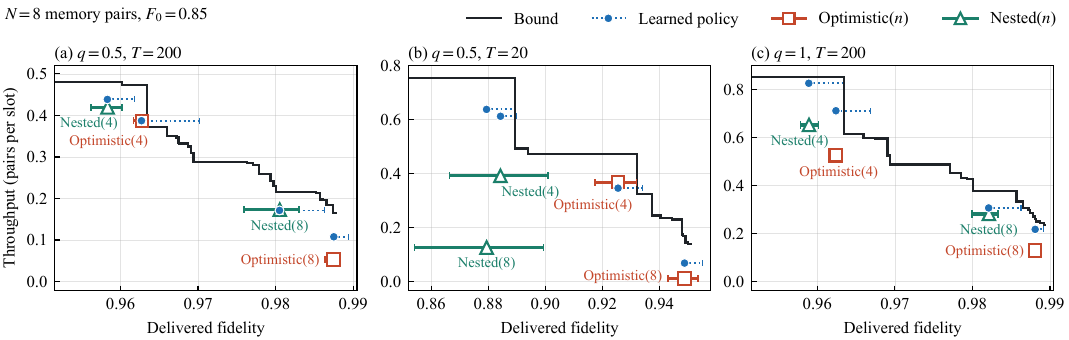}
  \caption{Throughput against delivered fidelity in three settings with $\numSlots = 8$ and $\Fzero = 0.85$. The line is the bound of Theorem~\ref{thm:bound} as a function of $\Fthr$; it ends at the largest $\Fthr$ for which the search for $\cstar(\Fthr)$ finished. Each baseline\protect\shortonly{, with $\alpha = 0.9$,} is drawn at the mean fidelity of its delivered pairs, all of which are counted, and its bar spans the $5\%$ and $95\%$ quantiles of their fidelities. Each point of the learned policy is a policy trained with $\Fthr$ equal to the mean fidelity of one baseline, so every pair it delivers has at least this fidelity; its dotted line ends at the $95\%$ quantile of the fidelities of its pairs. \protect\arxivonly{The baselines are shown for $n = 4$ and $8$ and $\alpha = 0.9$. With $n = 2$ they deliver a fidelity below the range shown, and a tree with $n = 16$ does not fit in $\numSlots = 8$ memory pairs; with $\alpha = 0.97$ their throughput is the same within $2\%$ in (a) and (c) and at most half as large in (b).}}
  \label{fig:frontier}
\end{figure*}

\section{Evaluation}
\label{sec:eval}

We compare the learned policy with the two baselines of Section~\ref{sec:baselines}, and all three with the bound of Theorem~\ref{thm:bound}, in two experiments.
The baselines are not an exhaustive set of heuristics for this problem; we use them to show how the best schedule changes with the setting, and that the learned policy adapts to it.
In the first experiment, $\Fthr$ is the fidelity that a baseline delivers, and we ask whether the learned policy reaches the throughput of that baseline.
In the second, $\Fthr$ is set by the application, and we ask how many pairs with fidelity at least $\Fthr$ each policy delivers as the parameters of the link and the number of memory pairs change.
In both, we also report the trees that the learned policy constructs, written as the strings of Section~\ref{sec:model-setting}, which show where it waits for outcomes but not the slots. Since the pairs that are available in a slot are random, the pairs that a policy delivers do not all have the same tree, and we report the share of the delivered pairs that have each string.

\subsection{Setup}
\label{sec:eval-setup}

A \emph{setting} is a choice of $\numSlots$, $\psrc$, $\Fzero$ and $\cohSlots$; the gate error is $\pgate = 0.005$ throughout.
We vary one parameter at a time, over $\numSlots \in \{4, 8, 16, 32, 64\}$, $\psrc \in \{1, 0.5, 0.2\}$, $\Fzero \in \{0.75, 0.80, 0.85, 0.90\}$ and $\cohSlots \in \{20, 60, 200, 600\}$, while the others are $\numSlots = 8$, $\psrc = 0.5$, $\Fzero = 0.85$ and $\cohSlots = 200$.
We report the throughput in pairs per slot.
\arxivonly{In this unit the results do not depend on the length of the link. For example, over $20$\,km of optical fiber, $\tcc = 0.1$\,ms, so $\cohSlots = 200$ is a coherence time of $20$\,ms and one pair per slot is $10\,000$ pairs per second. Over $100$\,km the same setting is a coherence time of $100$\,ms and $2000$ pairs per second.}

We run \Opt$(n,\alpha)$ and \Nest$(n,\alpha)$ for $n \in \{2, 4, 8, 16\}$ with $n \le \numSlots$, and for $\alpha \in \{0.9, 0.97\}$.
For the bound of Theorem~\ref{thm:bound} we find $\cstar(\Fthr)$ by brute force.
We compute the cost~\eqref{eq:cost} and the fidelity of every tree and schedule whose cost is below a limit, and we raise the limit until one of them reaches $\Fthr$; the cheapest such schedule gives $\cstar(\Fthr)$.

\shorter
We train every policy with PPO for $2 \times 10^{6}$ actions with discount factor $\disc = 0.99$, three times with different random seeds, and report the one with the highest throughput during training.
The encoder has $\numLayers = 2$ self-attention layers and embeddings of dimension $96$ for $\numSlots = 8$, and $\numLayers = 3$ and dimension $128$ for $\numSlots = 4$ and $16$; for $\numSlots = 32$ and $64$ we apply the policy trained for $\numSlots = 16$, without further training.
For every policy we then simulate $20$ independent runs of $20\,000$ slots each, after $1000$ warm-up slots, and report the mean throughput over the runs.

\subsection{Throughput at the Fidelities of the Baselines}
\label{sec:eval-frontier}

As defined in Section~\ref{sec:baselines}, a baseline delivers every pair that its tree produces, so the fidelity of its pairs is set by $n$ and the setting, not by the application.
This fidelity also varies from pair to pair, because the pairs in the tree of a delivered pair are retained for different numbers of slots.
Fig.~\ref{fig:frontier} shows what the baselines deliver in three settings.
The bars are narrow for $\cohSlots = 200$ and widen as the coherence time decreases, for \Nest$(4)$ from $0.004$ in (a) to $0.035$ in (b); \Opt{} has no spread for $\psrc = 1$, where every block has the same inputs.
We compare each baseline with the learned policy at the fidelity of that baseline: we train the policy with $\Fthr$ equal to the mean fidelity of the pairs that the baseline delivers.
The comparison favors the baseline, since we count all its pairs, including those with fidelity below $\Fthr$, whereas every pair of the learned policy has fidelity at least $\Fthr$.

In (a), \Nest$(4)$ delivers $87\%$ of the bound at its fidelity, $0.958$, so no policy can deliver over $15\%$ more. The learned policy delivers $5\%$ more, and for $65\%$ of its pairs it constructs $((\str{oo})[(\str{oo})])$, which waits for the outcomes of one first-level operation only.
It delivers the same throughput as \Opt$(4)$, $1\%$ less than \Nest$(8)$, and twice that of \Opt$(8)$.
In (b), where $\cohSlots$ is ten times smaller, the baseline closest to the bound is \Opt$(4)$, with $78\%$ of it at fidelity $0.926$.
Four elementary pairs reach this fidelity only in the optimistic tree, which the learned policy constructs for $70\%$ of its pairs; it delivers $6\%$ less than the baseline, $54\%$ of whose pairs are below $\Fthr$.
At the fidelity of \Nest$(4)$, $0.884$, three elementary pairs suffice: the tree $(\str{o}(\str{oo}))$, in which the output of an operation on two elementary pairs is distilled with a third one, is delivered with fidelity $0.889$.
A balanced tree has $n = 2^d$ elementary pairs, so neither baseline constructs it; the learned policy constructs it for $90\%$ of its pairs and delivers $56\%$ more than \Nest$(4)$.

In (c) neither baseline is close to the bound up to fidelity $0.963$, although the cheapest schedule there is the optimistic tree on four pairs.
\Opt$(4)$ delivers $62\%$ of the bound: in every slot it executes one block, on the three pairs retained from the previous slot and one new pair, and retains the other three new pairs.
\Nest$(4)$ delivers $77\%$; its tree waits for the outcomes of both first-level operations and costs $12\%$ more than the optimistic one.
The learned policy trained for the fidelity of \Nest$(4)$ delivers $97\%$ of the bound, $27\%$ more than \Nest$(4)$.
It applies the optimistic tree to four new pairs and, instead of retaining the pairs left over, applies an operation to them at once and completes $((\str{oo})[(\str{oo})])$ in the next slot.
At the fidelity of \Opt$(4)$, $0.9624$, the tree $((\str{oo})[(\str{oo})])$ no longer reaches $\Fthr$, since its pair has fidelity $0.9618$.
The learned policy then uses the pairs left over in a larger tree, $((\str{oo})[(\str{o}[(\str{oo})])])$, and still delivers $35\%$ more than the baseline.
At the fidelity that a baseline delivers, the learned policy thus reaches the throughput of that baseline where the baseline is close to the bound, and exceeds it elsewhere.

\begin{table}[b!]
  \caption{Throughput of Pairs With Fidelity at Least $\Fthr = 0.95$, in Pairs per Slot}
  \label{tab:qos}
  \centering
  \begin{tabular}{@{}lccccr@{}}
    \toprule
    Setting & Bound & \Opt & \Nest & Learned & Gain \\
    \midrule
    $\numSlots = 4$ & 0.240 & 0.134 & 0.177 & 0.195 & $+10\%$ \\
    $\numSlots = 8$ & 0.480 & 0.387 & 0.419 & 0.439 & $+5\%$ \\
    $\numSlots = 16$ & 0.960 & 0.855 & 0.899 & 0.912 & $+1.5\%$ \\
    $\numSlots = 32^{\dagger}$ & 1.921 & 1.797 & 1.860 & 1.873 & $+0.7\%$ \\
    $\numSlots = 64^{\dagger}$ & 3.842 & 3.695 & 3.779 & 3.794 & $+0.4\%$ \\
    \addlinespace[3pt]
    $\psrc = 1$ & 0.852 & 0.526 & 0.653 & 0.826 & $+27\%$ \\
    $\psrc = 0.2$ & 0.228 & 0.161 & 0.195 & 0.195 & $0.0\%$ \\
    \addlinespace[3pt]
    $\cohSlots = 60$ & 0.473 & 0.365 & 0.231 & 0.367 & $+0.5\%$ \\
    $\cohSlots = 600$ & 0.481 & 0.388 & 0.421 & 0.439 & $+4\%$ \\
    \addlinespace[3pt]
    $\Fzero = 0.75$ & 0.147 & 0.020 & 0.083 & 0.098 & $+17\%$ \\
    $\Fzero = 0.80$ & 0.241 & 0.034 & 0.206 & 0.210 & $+1.9\%$ \\
    $\Fzero = 0.90$ & 0.598 & 0.476 & 0.479 & 0.556 & $+16\%$ \\
    \bottomrule
  \end{tabular}

  \vspace{3pt}
  \begin{minipage}{\columnwidth}
    \footnotesize
    In each row one parameter has the value given; the others are $\numSlots = 8$, $\psrc = 0.5$, $\cohSlots = 200$ and $\Fzero = 0.85$. \Nest{} is extended to use $\Fthr$ (see text). Each baseline is reported with the $n$ and $\alpha$ that give its highest value in that row. Gain: learned policy relative to the better baseline. The standard error is below $1\%$ of every value. $^{\dagger}$Learned policy trained for $\numSlots = 16$.
  \end{minipage}
\end{table}

\subsection{Throughput at a Required Fidelity}
\label{sec:eval-required}

\shorter
Fig.~\ref{fig:frontier} compares the policies at the fidelities that the baselines deliver.
These are few: a baseline of Section~\ref{sec:baselines} delivers one range of fidelities for each $n$, so an application whose threshold lies between two of them must use the larger tree.
In the second experiment we therefore fix the threshold, as an application does, with $\Fthr = 0.95$ and $\Fthr = 0.97$ as two examples, and count for every policy only the delivered pairs with fidelity at least $\Fthr$, as in~\eqref{eq:optimal-throughput}.
Since a pair below $\Fthr$ is not counted, we extend \Nest{} for this experiment so that it does not deliver such a pair.
When a pair of level $d$ becomes available with fidelity below $\Fthr$, the policy reduces its level to $d-1$ and combines it with a pair of that level, and it repeats this until the fidelity is at least $\Fthr$. The tree of a delivered pair can therefore end up with more than $n$ elementary pairs.
\Opt{} meets $\Fthr$ through the choice of $n$ and $\alpha$ alone.
The fidelity of its pairs varies less than under \Nest{}, as the bars of Fig.~\ref{fig:frontier} show.
Table~\ref{tab:qos} reports the throughput at $\Fthr = 0.95$ as the setting changes, and shows in which settings one of the two baselines suffices: the learned policy delivers as much as the better one where it fits the threshold, and more where neither does.
For $\cohSlots = 60$, four elementary pairs reach $0.95$ only in the optimistic tree, and only if they are retained briefly. \Opt$(4)$ with $\alpha = 0.97$ then delivers $77\%$ of the bound, and the learned policy, which constructs this tree for $93\%$ of its pairs and otherwise adapts by constructing a larger one, $0.5\%$ more.
For $\Fzero = 0.80$, in contrast, four elementary pairs no longer reach $0.95$, and the cheapest schedule, which gives $\cstar(\Fthr)$, is $([(\str{oo})][([(\str{oo})][(\str{oo})])])$, on six elementary pairs with every operation waiting for the outcomes of its inputs.
\Nest$(4)$ constructs this tree for $94\%$ of its pairs and the learned policy for $99.98\%$; the learned policy delivers $1.9\%$ more.
The gain is largest for $\psrc = 1$, $27\%$, as in Fig.~\ref{fig:frontier}(c), and it is $16\%$ for $\Fzero = 0.90$.
For $\Fzero = 0.75$ it is $17\%$: a tree of eight elementary pairs reaches $0.95$ only if its pairs are retained briefly.
The learned policy adapts by not waiting for every outcome and achieves this for $67\%$ of its pairs, against $39\%$ for \Nest$(8)$; the other pairs are distilled further.

The gain also depends on $\numSlots$.
In the five rows that vary $\numSlots$, the cheapest schedule is the tree of \Nest$(4)$, so this baseline falls short of the bound only through the few pairs left without a partner in a slot. These become a smaller share of all pairs as $\numSlots$ grows.
\Nest$(4)$ delivers $74\%$ of the bound for $\numSlots = 4$ and $98\%$ for $\numSlots = 64$, and the gain of the learned policy falls from $10\%$ to $0.4\%$.
To test how a learned policy generalizes to a larger memory, we applied the policy trained for $\numSlots = 16$ to $\numSlots = 32$ and $64$ without further training; for $\Fthr = 0.95$ it delivers $98\%$ and $99\%$ of the bound.
For $\Fthr = 0.97$ (not in Table~\ref{tab:qos}), in contrast, the policy trained for $\numSlots = 16$ delivers $94\%$ of the bound for $\numSlots = 16$, $3.6\%$ more than \Nest$(4)$, but $87\%$ for $\numSlots = 32$ and $64$, $8\%$ and $10\%$ less than \Nest$(4)$.
A learned policy can thus be applied to a larger memory without retraining, but it need not remain close to the bound. A larger memory can favor schedules that did not arise in training, and the policy may then have to be retrained.

\section{Conclusion}
We derived an upper bound on the throughput of every policy for online distillation scheduling and trained, with RL, a policy that decides in every slot which pairs to distill and, for each operation, whether to wait for its outcomes.
Its optimism depends on the setting: it waits for almost no outcomes when memories decohere fast, for nearly all of them when high fidelity requires large trees, and in other settings for the outcomes of some operations but not of others.
Compared with two baselines, one that always waits for the outcomes and one that never does, the learned policy delivers, in the settings we tested, as many pairs as the better one where one of them suits the setting, and more where neither does.
Where a baseline is close to the bound, no policy can deliver much more, so the bound shows where training pays off.
Our model can be adapted to other distillation protocols, for example the optimized circuits of~\cite{krastanov2019}, some of which take more than two input pairs.
It assumes that local operations are instantaneous and that any two stored pairs can be distilled together; relaxing these assumptions for specific hardware is left for future work.
\bibliographystyle{IEEEtran}
\bibliography{references}

\end{document}